\documentclass{aa}
   
\usepackage{graphicx,epsfig,fancyhdr,rotating,amsmath,natbib}   
\begin{document}   
\title{Dynamics and evolution of an eruptive flare}   
\author{L. Teriaca\inst{1}, A. Falchi\inst{2}, R. Falciani\inst{3}, 
	G. Cauzzi\inst{2} \and L. Maltagliati\inst{1,2}} 
 
\offprints{L. Teriaca,\\ \email{teriaca@linmpi.mpg.de}} 
 
\institute{Max-Planck-Institut f\"{u}r Sonnensystemforschung,    
        Max-Plank Str. 2, 37191 Katlenburg-Lindau, Germany 
\and 
INAF-Osservatorio Astrofisico di Arcetri,  
	Largo Enrico Fermi 5, 50125 Firenze, Italy 
\and 
Dipartimento di Astronomia e Scienza dello Spazio, Universit\`{a} di 
	Firenze, Largo Fermi 2, 50125 Firenze, Italy} 
 
\date{Received 22 February 2006 / Accepted 23 May 2006} 
   
\abstract 
{} 
{We study the dynamics and the evolution of a C2.3 two-ribbon flare, 
developed on 2002 August 11, 
during the impulsive phase as well as during the long gradual phase.  
To this end we obtained multiwavelength observations 
using the CDS spectrometer 
aboard SOHO, facilities at the National Solar 
Observatory/Sacramento Peak, and the TRACE and RHESSI spacecrafts.} 
{CDS spectroheliograms in the Fe~{\sc xix}, Fe~{\sc xvi}, O~{\sc v} and 
He~{\sc i} lines allows us to determine the velocity field at different
heights/temperatures 
during the flare and to compare them with the chromospheric velocity fields 
deduced from H$\alpha$ image differences. TRACE images in the 17.1~nm band 
greatly help in determining the morphology and the evolution of the flaring 
structures.} 
{During the impulsive phase a strong 
blue-shifted Fe~{\sc xix} component ($-200$~km~s$^{-1}$) is observed at the 
footpoints of the flaring loop system, 
together with a red-shifted emission of O~{\sc v} and He~{\sc i} lines 
($20$~km~s$^{-1}$). 
In one footpoint simultaneous H$\alpha$ data are also 
available and we find, at the same time and location, downflows with an inferred 
velocity between 4 and 10~km~s$^{-1}$.  
We also verify that the ``instantaneous'' momenta of the oppositely 
directed flows detected in Fe~{\sc xix} and H$\alpha$ are equal within one 
order of magnitude. These signatures are in general agreement 
with the scenario of explosive chromospheric evaporation. 
Combining RHESSI and CDS data after the coronal upflows have ceased, 
we prove that, independently from the filling factor, 
an essential contribution to the density of the post-flare loop 
system is supplied from evaporated chromospheric material. 
Finally, we consider the cooling of this loop system, that 
becomes successively visible in progressively colder 
signatures during the gradual phase. We show that the observed 
cooling behaviour can be obtained    
assuming a coronal filling factor of $\approx$~0.2~to~0.5.} 
{} 
\keywords{Sun: flares -- Sun: chromosphere -- Sun: corona -- Line: profiles} 
   
\authorrunning{Teriaca et al.} 
%\titlerunning{Dynamics and evolution of an eruptive flare} 
   
\maketitle 
   
%%%%%%%%%%%%%%%%%%%%%%%%%%%%%%%%%%%%%%%%%%%%%%%%%%%%%%%%%%%%%%%%%%%%%%%%%%%%%%%   
%-----------------------------------   
%   
\section{Introduction}\label{intro}   

Most solar flare models are based on the idea that flares arise from the   
sudden release of free magnetic energy stored in the corona in   
non-potential configurations \citep[e.g.,][]{Klimchuk-etal:88}. The bulk   
of the energy is released through  reconnection of the magnetic field   
lines in the  corona \citep[]{Petschek:64, Kopp:76, Shibata:96} and then   
transported down to the chromosphere along the magnetic field lines by    
accelerated particles    
or by a thermal conduction front.   
Hydrodynamic simulations indicate that both non-thermal electrons   
\citep[]{Fisher-etal:85a, Fisher-etal:85b, Mariska-etal:89} and a thermal conduction   
front \citep[]{Gan-etal:91} should lead to the evaporation of   
chromospheric material.  
%Its overpressure drives upward   
%motions in the corona and downward motions in the chromosphere and it is   
%expected that the momenta of the oppositely moving plasma should be balanced.  
%   
Models by \citet{Fisher-etal:85a} result in upflows (gentle evaporation) or 
downflows (explosive evaporation), on both transition-region (TR) 
and chromosphere, depending  on the energy flux of the 
impinging electrons, with downflows predicted for large energy fluxes.  
In the case of explosive evaporation the overpressure of evaporated 
plasma drives upward 
motions in the corona and downward motions in the chromosphere and it is 
expected that the momenta of the oppositely moving plasma should be balanced. 
The chromospheric evaporation   
injects mass into the corona to  fill coronal loops that appear as   
the soft X-ray (SXR) flare and  that eventually   
cool down radiatively and/or conductively forming cold H$\alpha$ post-flare   
loops \citep[see, e.g.,][]{Moore-etal:80}. Although this general picture   
seems well established, the details of some of the key processes assumed   
to take place remain to be verified. The small temporal and spatial scales   
involved, together with the need for comprehensive temporal and height   
coverage, conjure up to make ``complete''  flare observations  a very   
demanding task.

Blue-shifts 
of lines formed at flare temperatures %($\approx10^7$~K) 
during the impulsive phase  
(corresponding to upflow velocities in excess of several hundreds  
kilometre per second) 
have been observed since the early 1980s    
\citep[e.g.,][]{Doschek-etal:80, Antonucci-etal:82} and confirmed 
with the Bragg Crystal Spectrometer aboard Yohkoh   
\citep[BCS,][]{Culhane-etal:91}. However, the majority of these   
observations revealed only a blue asymmetry,    
 indicating the presence of a large, static component   
 even in the early phases of the flare 
\citep[e.g.,][]{Bentley:94, Gan:97}, rather than wholly blue-shifted   
profiles as predicted by hydrodynamic models    
\citep[e.g.,][]{Emslie:92, Gan:95}.    
To which extent the lack of spatial resolution might have influenced   
these observations is not easily quantifiable (neither BCS nor    
earlier spectrometers provided any effective spatial resolution    
on the solar surface).    
Since reconnection is expected to occur at very small   
spatial scales, it might well be that the associated chromospheric evaporation   
takes place in rather confined   
portions of the solar surface. This is indeed the case in eruptive flares,   
for which, at any given time, the evaporation in the reconnection scenario   
occurs in a very thin shell ($\le$~1$''$)  located on the outer part of the   
flaring loop system \citep[]{Forbes:96,Falchi:97}, and the heating of  
individual field lines is supposed to last only 100~s or so   
\citep[]{Forbes:03}. The smearing introduced by observing over the whole   
disk would then significantly alter the relevant signals.   
Hydrodynamic simulations modelling solar flare    
emission with multiple  loops \citep[]{Hori:98,Warren:05}    
have shown that it is possible to   
reproduce the line profiles dominated by the stationary component (as   
observed by BCS) with a simple model based on the successive independent   
heating of   
small scale threads. Only the simulated line profiles for individual   
threads do show strongly blue shifted line profiles.   
     
Spatially and temporally    
resolved observations of flare lines have finally become    
available with the advent of the Coronal Diagnostic Spectrometer 
\citep[CDS,][]{Harrison-etal:95} aboard SOHO, which provides stigmatic slit 
images in the 31 to 38~nm (NIS-1) and 51 to 63~nm (NIS-2) range with a spatial   
resolution of few seconds of arc and a cadence depending upon the adopted observing   
program. Various authors have since reported the presence   
of blueshifted emission in hot coronal lines during flares,    
signalling upflows of up to $\approx200$~km~s$^{-1}$   
\citep{Czaykowska-etal:99,DelZanna-etal:02b,    
Brosius:03,Teriaca-etal:03,Brosius:04}, but   
the total number of events analysed up to now   
remains small. Moreover, the same authors report about    
intricate patterns of   
blue- and red-shifted emission in TR lines, depending both   
on the location within the flaring region and the phase of the flare, that   
still need to be consistently explained within the theory of chromospheric   
evaporation. Finally, spatially and temporally resolved observations  
of flare lines 
need to be combined with simultaneous chromospheric   
observations, in order to verify the spatial and temporal 
coincidence of chromospheric downflows with hot plasma upflows. 
In fact, while   
many authors have reported on the simultaneous appearance 
of redshifts in chromospheric lines and blue-shifts in flare 
lines during the early phase of flares  \citep[e.g.,][]   
{Canfield-etal:90b, Falchi:92,   
Wulser-etal:94}, to date the only direct proof that such shifts   
%lines (Fe~{\sc xix})    
arise from spatially coincident areas has been provided by   
\citet{Teriaca-etal:03}. However, due to the  observational setup,    
their coronal and chromospheric  data were not simultaneous, 
leading to the need of further observations.   
 
In this context, we report here on comprehensive,  multi-instrument   
observations of a small eruptive flare, covering its temporal 
evolution  from the impulsive phase and related evaporation   
throughout the gradual phase and appearance of cold postflare loops.   
Spatially resolved CDS spectra allow us to study the dynamics of the upper   
atmosphere at various times during the flare, while simultaneous    
chromospheric observations provide a test for the chromospheric   
evaporation scenario. RHESSI data acquired during the gradual phase, combined  
with CDS measurements, provide a final element in support of the chromospheric  
evaporation.  
Other SOHO instruments and TRACE, provide   
various context information and help us in defining and constraining the   
flare evolution. 
     
%===================================================================   
\begin{figure}[!t] 
\resizebox{\hsize}{!}{\includegraphics{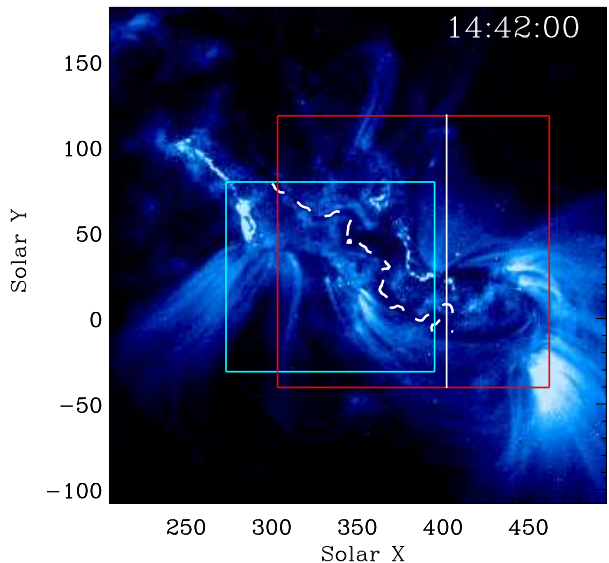}} 
  %\resizebox{\hsize}{!}{\includegraphics{5065f1_c.eps}}   
\caption{TRACE image (logarithmic scale) in the 17.1~nm    
band showing AR~10061 at flare peak. The dark grey (red in the electronic   
version) box indicates the UBF FOV while the area rastered by CDS/NIS is    
represented by the light grey (blue) box. The white vertical line marks the    
position of the DST spectrograph slit. Here and in the    
following images, the white dashed line shows the apparent magnetic    
inversion line obtained from MDI data in the region around the filament.}   
\label{fig:ctx}   
\end{figure}   
%%===================================================================   
   
\section{Observations and data analysis}\label{obs}   
The observations were acquired during a   
coordinated campaign between ground based    
and SOHO instruments aimed at studying flare events    
by sampling the solar atmosphere from the    
chromosphere to the corona. The main instruments    
involved were the CDS/NIS spectrometer aboard    
SOHO and facilities at the NSO/Sacramento Peak. The analysed flare 
developed at about 14:40~UTC in region NOAA~10061 on 2002 August 11   
(N10W20). Although    
small (GOES class C2.3), the flare was clearly eruptive in character with   
an impulsive rise and a long decay phase, and developed two    
ribbons visible at various wavelengths on opposite sides of the magnetic   
neutral line. A second flare, not studied here, developed   
in the same region around 16:25~UTC. 
   
\subsection{Data}\label{data} 
{\it Ground-based data} -- Monochromatic images at several    
wavelengths    
have been acquired, with a temporal cadence of a few seconds,   
by the tunable Universal    
Birefringent Filter (UBF: H$\alpha$ line centre, $-60$~pm,   
$+60$~pm and $-150$~pm off centre; He~{\sc i}~D3; Na~{\sc i}~D2 and in    
the continuum) and by the Zeiss filter (H$\alpha$ $+150$ pm) at    
the Dunn Solar Telescope of the National Solar    
Observatory (DST/NSO). The field of    
view (FOV) was of about $150'' \times150''$ with a spatial    
scale of  $0.5''\times0.5''$. Spectra    
have been acquired with the Horizontal    
Spectrograph (HSG) in three chromospheric lines (Ca~{\sc ii}~K,   
H$\gamma$ and He~{\sc i}~D3) with a temporal cadence of 4~s and the    
fixed slit positioned as indicated in Fig.~\ref{fig:ctx}.   
   
{\it Space-born data} -- Spectroheliograms of a $120''\times110''$ area    
(mostly overlapping the UBF FOV, see Fig.~\ref{fig:ctx}) were    
obtained with the Normal Incidence Spectrometer    
(NIS) of the CDS experiment \citep{Harrison-etal:95}    
starting at 14:26:27 UTC. Each raster was obtained in    
$\approx210$~s by stepping the $4''$   
wide slit eastward in 19 steps of $6''$. On board binning over two rows   
yielded a pixel size of 3.4$''$ along the slit. Spectra were    
acquired in the He~{\sc i}~58.43~nm ($2\times10^{4}$~K),   
O~{\sc v} 62.97~nm ($2.5\times10^{5}$~K),   
Fe~{\sc xvi} 36.08~nm ($2\times10^{6}$~K) and   
Fe~{\sc xix} 59.22~nm ($8\times10^{6}$~K) lines. Every ten rasters    
an additional    
step of $\approx 5''$ westward was performed to compensate for    
the solar rotation.   
During the entire period of CDS observations AR~10061    
was constantly monitored by the Transition Region and    
Coronal Explorer \citep[TRACE,][]{Handy-etal:99} that    
provided images in the 17.1 nm band (Fe~{\sc ix -- x}, $\approx 10^{6}$~K)    
with a spatial scale of $1''$ pixel$^{-1}$ (on board $2\times2$    
binning) and a cadence of $\approx7.5$~s.   
The Reuven Ramaty High-Energy    
Solar Spectroscopic Imager \citep[RHESSI,][]{Lin-etal:02} exited 
its night  
%exited  from the Earth shadow  
a few minutes after the peak phase of the    
flare, and provided  information on     
hot ($\approx 10^{7}$~K) thermal flare plasma as well as on non-thermal   
components.
From RHESSI data in the energy band 3~to~10~keV, we derived images 
with spatial resolution up to $2''$ and   
spectra of the spatially integrated flux with 1~keV    
spectral resolution. 
Unfortunately, due to particle precipitation events, only three different   
times could be analysed during the gradual phase of the flare.   
Finally, MDI \citep{Scherrer-etal:95} full-disk magnetograms    
taken at 14:27:34~UTC and 16:03:34~UTC were used to    
determine the position of the photospheric apparent magnetic inversion line.   
 
\subsection{Data co-alignment}\label{coal}   
Data from the different instruments were aligned using    
MDI continuum data as a reference. UBF continuum    
images (and, hence, all DST data) were easily    
aligned first. Afterwards, CDS images in the He~{\sc i} 58.4~nm line    
were aligned with the UBF images acquired in H$\alpha$ line centre.    
Moreover, the flare kernels at the loop footpoints in the    
TRACE 17.1 nm images appear strikingly similar to    
those observed in H$\alpha$ line centre, allowing a very good    
relative alignment. The brightening of the    
flare kernels is also very clear in the CDS O~{\sc v} images,    
providing further constraints. We estimate the    
alignment between TRACE, MDI and ground data to    
be precise within $1''$ and within $2''$ with CDS. RHESSI    
images are more difficult to align and the precision is    
estimated around $5''$. Each spectrum or image has been    
tagged with its precise acquisition time so that light and    
velocity curves obtained using data from different    
instruments can be properly computed.    
 
\subsection{Measurement of line-of-sight motions}\label{speed}    
{\it Ground-based data} -- Strong flows in the flaring chromosphere are often   
measured through spectral line asymmetries, defined generally with the   
bisector \citep[]{Zarro-etal:88, Falchi:92}.   In the portion of the   
flare ribbon intersected by the HSG slit,  the Ca~{\sc ii}~K line shows   
indeed an emission core with a red asymmetry. The H$\gamma$ line, instead,   
remains in absorption (with a central radiance higher than in the   
reference quiet area), so that several blends in its wings prevent a   
reliable measure of the otherwise visible asymmetry. The He~{\sc i}~D3   
line is only barely visible. The velocity component along the line of   
sight (LOS) for the HSG spectra was hence computed using the bisector of   
the emission core of the Ca~{\sc ii}~K line, in all 16 pixels in the   
ribbon area and for 41 different times.  
%In Fig. \ref{fig:profilo_caii}   
%the spectrum acquired in a flaring kernel is shown together with the bisector,   
%clearly outlining the red asymmetry of the line, in this case resulting   
%in a maximum velocity of 12~km~s$^{-1}$.   
Unfortunately, the (fixed) spectrograph slit was positioned slightly   
outside the FOV covered by the CDS rasters (see Fig.~\ref{fig:ctx}), so   
that no direct comparison of velocities deduced from HSG spectra can be   
performed. However, in the area of the flaring ribbons within the CDS FOV,   
we could use the H$\alpha$ images acquired at several wavelengths to   
estimate the chromospheric velocities. To this end,  
difference images at $\pm$60~pm and $\pm$150~pm were computed.  
In all flaring kernels the differences 
between the red and  blue wings of   
H$\alpha$ were positive and very close in value,  
while the radiances in both the centre and in    
the wings remained higher than in the reference quiet area. This implies  
an asymmetric H$\alpha$ profile with a   
stronger emission in the red part of the line  
and hence a downward velocity for all flaring kernels. 
To estimate the value of 
this velocity, we calibrated 
the radiance differences obtained from the H$\alpha$ images 
in the kernel where HSG spectra have been acquired 
with the velocities obtained from the Ca~{\sc ii}~K spectra.  
This was done for 
all times and pixels for which a reliable estimate of the  
Ca~{\sc ii}~K 
velocity was available (about 500 points).  
We find that the average H$\alpha$ radiance 
difference observed in the flare kernels, both at $\Delta\lambda \pm 
60$~pm and $\Delta\lambda \pm 150$~pm, corresponds to a downward velocity 
between 4 and 10~km~s$^{-1}$. 
Similar velocity values have been found from H$\alpha$ line profiles in 
the case of small (GOES class B~--~C) flares \citep[e.g.,][]{Schmieder:98}. 
 
{\it CDS spectra} -- Spectra acquired after recovery of the SOHO   
spacecraft are characterised by broad and asymmetric line profiles   
that were fitted with an opportune template provided within   
the CDS software\footnote{CDS Software Note 53 by W.T. Thompson at   
{\tt http://solar.bnsc.rl.ac.uk/software/notes.shtml}.}.   
The template consists of a    
Gaussian component (with all parameters free to   
vary) and a wing (non-Gaussian)   
component whose parameters are established for both   
NIS-1 and NIS-2 spectra.   
 
The analysis of line positions reveals that no trends (either parallel or   
perpendicular to the slit direction) are present within a single raster.   
However, the analysis of the raster-averaged NIS-2 line positions versus time   
reveal a small linear trend that is equal in He~{\sc i} and O~{\sc v}.    
The shift amounts to $\approx$~20~km~s$^{-1}$ over two hours. This   
trend has been accounted for by providing a corrected (shifted) wavelength   
vector for each NIS-2 line in each raster. No such a trend is visible in the   
Fe~{\sc xvi} NIS-1 line.
     
Flows in the He~{\sc i}~58.43~nm, O~{\sc v}~62.97~nm, and Fe~{\sc xvi}~36.08~nm   
lines have been measured fitting a single component to the line profiles.    
These lines have an obvious pre-flare component, so that a reference wavelength   
has been obtained averaging the central positions derived from the fitting   
for the whole dataset.   
Velocities have then been computed simply using the Doppler-shift of the flare   
profiles with respect to the reference wavelength.  
%Since both He~{\sc   
%i}~58.43 nm and O~{\sc v}~62.97~nm provide essentially the same   
%information, in the following we will refer mainly to the latter.   
   
The measurement of flows in the hot ($\log~T/$K=6.9) Fe~{\sc xix} 59.22~nm line   
requires a more careful analysis. The line appears only during flares,   
so great caution is needed in choosing the reference profile    
\citep[see discussions in][]{Teriaca-etal:03,Brosius:03}.   
The line profiles obtained averaging all CDS pixels with a radiance    
greater than 0.16~W~m$^{-2}$~sr$^{-1}$ are identical in width and position   
in three successive scans during the late phase of the flare    
(from 14:53:10 to 15:00:10~UTC, times at raster centres),   
so their average was used as our   
reference profile.   
A multi-component fitting was hence applied to the profiles in various   
pixels and times,   
constraining the rest component by imposing width and position of the   
reference profile, while the background \citep[characterised by the
Fe~{\sc xii} 59.26 nm line, see][]{DelZanna:05} 
was constrained through the average non-flare   
spectrum.    
%   
%===================================================================   
\begin{figure}[!t] 
\resizebox{\hsize}{!}{\includegraphics{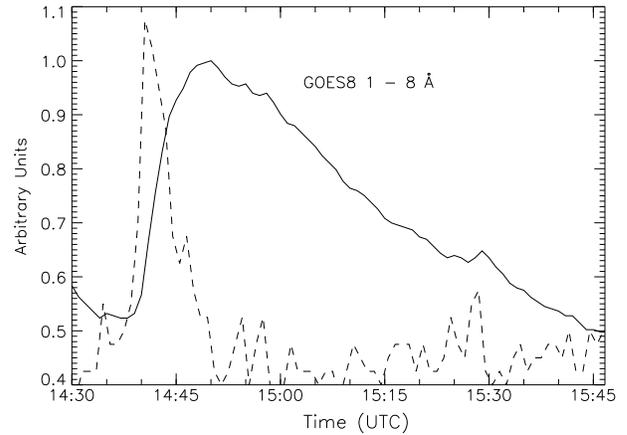}}  
   \caption{Soft X-ray full disk flux at 1~UA in the 1~to~8~\AA~band   
   around the time of the flare (GOES~8 data). The time derivative   
   (dashed line) represents a proxy for the hard X-ray curve (Neupert   
   effect) and shows its maximum around 14:41:20~UTC.   
   }   
   \label{fig:sxr}   
\end{figure}   
%===================================================================   
%  
Uncertainties on best-fit parameters are evaluated accounting for data noise   
(photon statistics,  pulse height distribution and readout noise) and the   
average 3-$\sigma$ uncertainty amounts to 15~to~18~km~s$^{-1}$ over the active 
region for single component fitting and to 25~to~30~km~s$^{-1}$ for double 
component fitting.    
   
\section{Flare evolution }\label{flareev}   
%===================================================================   
\begin{figure*}[!ht] 
\sidecaption 
\includegraphics[width=12cm]{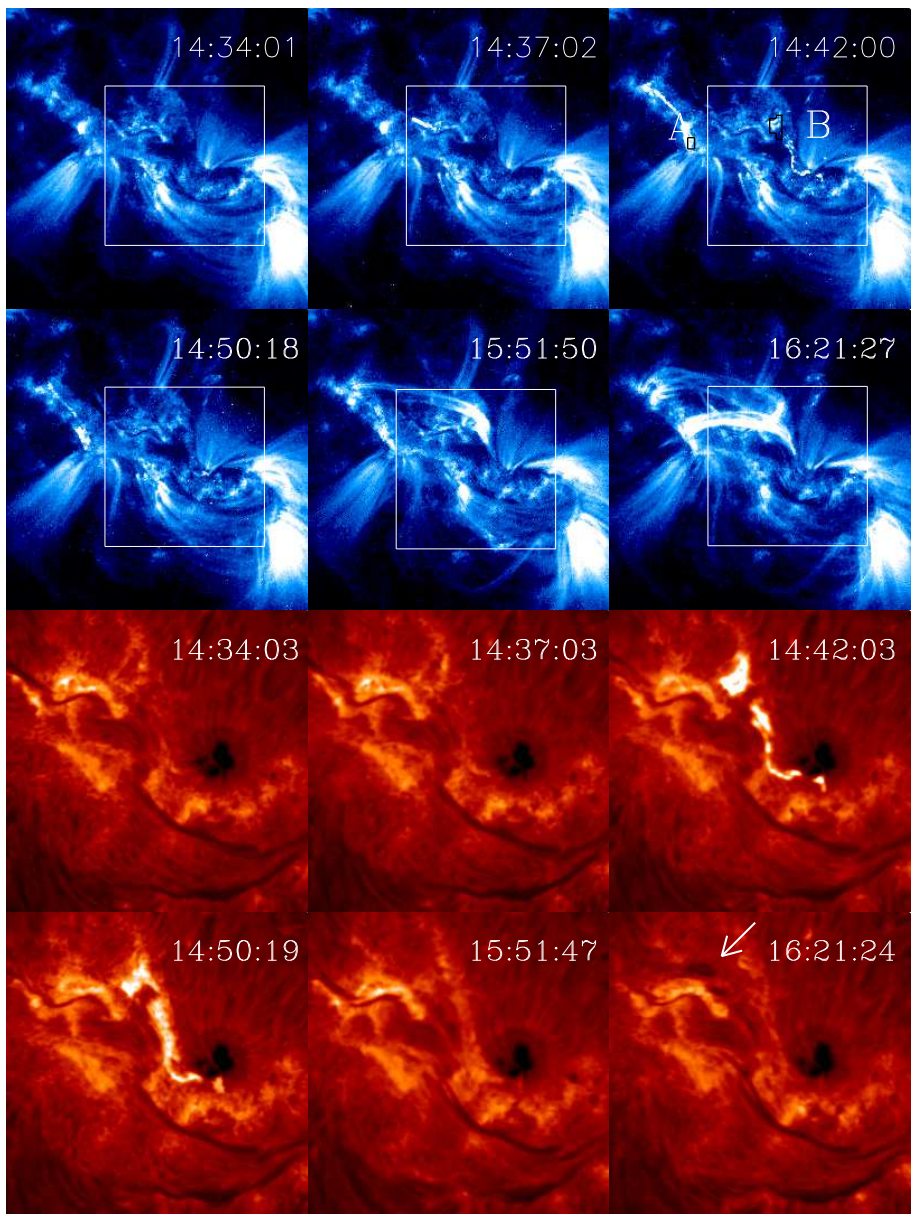}
  \caption{Evolution of the flare: TRACE images (logarithmic scale) in the   
  17.1~nm band are shown in the first two rows and H$\alpha$ images in 
  the last two rows. The white box in TRACE images indicate   
  the H$\alpha$ FOV.    
  The images at 14:34~UTC show the region just before the flare.   
  The areas (labelled A and B) over which data were integrated to obtain the   
  light and velocity curves shown in Fig.~\ref{fig:lgt2} are indicated in   
  black on the TRACE image taken at 14:42:00~UTC.    
  The arrow in the last H$\alpha$ image indicates the cool   
  post-flare loops.}   
  \label{fig:evol}   
\end{figure*}   
%===================================================================   
   
Since RHESSI was in the Earth shadow during the initial phase, no hard 
X-ray observations were available   
to define the impulsive phase of the flare. Moreover, the low temporal    
resolution attained with the CDS rastering mode   
prevents the use of EUV radiance peaks as a proxy for hard X-ray (HXR)   
bursts, as was done, e.g., by    
\citet[]{Brosius:03}. Hence we   
used the time-derivative of the GOES SXR flux    
as a proxy for the HXR emission \citep[Neupert effect;][]{Neupert:68}.   
Both the 0.1~to~0.8~nm (1.6 to 12.4~keV) flux and its derivative are   
displayed   
around the time of   
the flare development in Figure~\ref{fig:sxr}.   
The main episode of energy release indicated by the Neupert peak    
is around 14:41:20~UTC, but significant   
signal is present up to the GOES maximum, i.e., up to around 14:48~UTC.   
We assume the gradual phase starts after this time.   
   
Figure~\ref{fig:evol} outlines the main phases of the flare evolution   
as seen in the TRACE 17.1~nm channel and H$\alpha$ centre.    
A filament located on the magnetic neutral line is clearly visible    
in both spectral ranges before the flare.    
The activation of a portion of this filament is observed     
around 14:13~UTC in the H$\alpha$ images    
and lasts up to    
14:37~UTC, when it becomes clearly visible also in the TRACE    
images. The eruption was detected by CDS in the    
O~{\sc v} line, where upward speeds around 30~km~s$^{-1}$ were    
detected at 14:36:33~UTC.    
   
Few minutes later the   
footpoints of a large loop system (spanning over $6\times 10^4$~km 
on the solar surface) start to    
brighten, reaching their maximum around 14:42~UTC.    
The flare ribbons are clearly visible and nearly identical in    
both TRACE and H$\alpha$ images. Although the east-most ribbon is   
outside of the FOV of the UBF H$\alpha$ images, we verified its   
existence in the images acquired by the HASTA telescope    
\citep[see Fig.~1 in][]{Maltagliati-etal:06}.    
During the impulsive phase of the flare, the ribbons are associated with    
large coronal upflows and chromospheric and TR downflows    
(see next section). The brightenings last several    
minutes and fade off around 14:50~UTC in the 17.1~nm    
band while their decay is much longer in H$\alpha$.    
%===================================================================   
\begin{figure*}[!ht] 
\sidecaption 
\includegraphics[width=12cm]{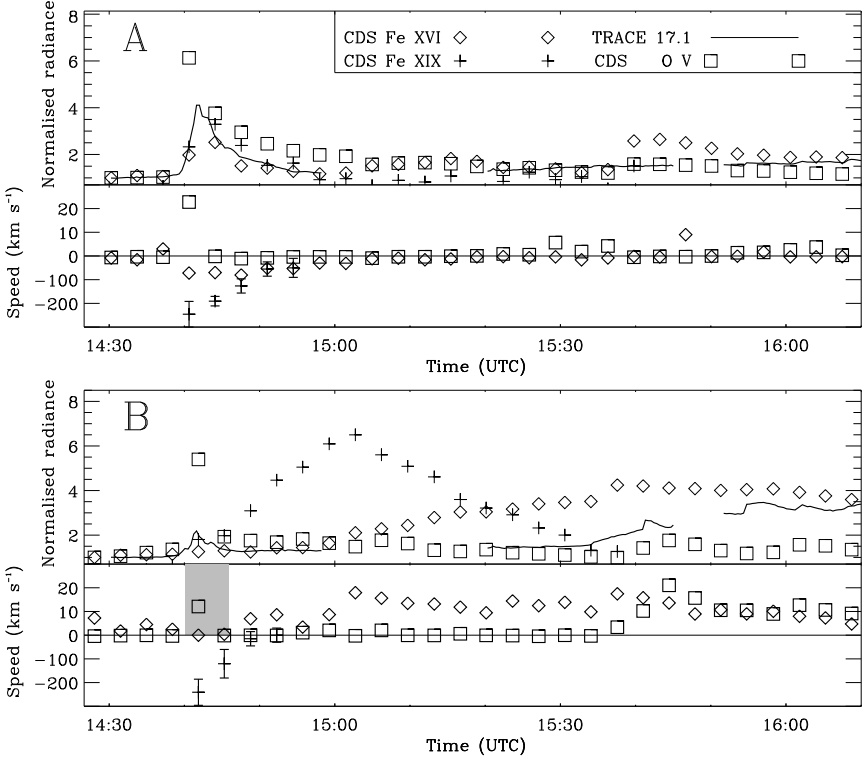}   
   \caption{Light and velocity curves at the footpoints (areas A and   
   B in Fig.~\ref{fig:evol}) of the loop system.   
   For clarity reasons uncertainties  
   are only shown for the Fe~{\sc xix} speeds. O~{\sc v}, Fe~{\sc xvi}, and   
   TRACE radiances are normalised to pre-flare values while Fe~{\sc xix}   
   radiances are normalised to the average value reached after the flare   
   (around 15:32~UTC). The average   
   uncertainties in the normalised radiance are of about 0.15.  
   On footpoint A, the average uncertainties in 
   speed are 12~km~s$^{-1}$ (O~{\sc v}) and 20~km~s$^{-1}$ (Fe~{\sc xvi}).  
   On footpoint B, the average 
   uncertainties in speed are 9 and 12~km~s$^{-1}$ for O~{\sc v} and 
   Fe~{\sc xvi}.   
   The shaded area represents the time during which downflows 
   between 4 and 10~km~s$^{-1}$ were observed in H$\alpha$. (We remind that 
   only for footpoint B we have chromospheric observations). 
   On both panels, the scale of downward (positive) velocities 
   has been expanded for sake of clarity.} 
   \label{fig:lgt2}   
\end{figure*} 
%===================================================================   
 
Immediately after the GOES maximum, the first     
RHESSI spectrum (14:50~UTC) between 4 and 10~keV  
shows a non-thermal component indicating    
that electron acceleration is still present. During the gradual phase   
of the flare, two cooling loop systems connecting the ribbons   
appear in the TRACE images, one around 15:40~UTC, and the other   
around 16:03~UTC. Fig.~\ref{fig:evol} shows them at the time of maximum   
radiance.   
Finally, these loops appear also as dark    
features in the H$\alpha$ images at 16:21~UTC, revealing that the plasma has    
cooled down to temperatures around $10^4$~K.    
 
In the following, we will concentrate on the footpoints of the second   
loop system, contained in the CDS FOV,    
where most of the brightenings   
and velocity episodes are visible in our dataset.    
Figure \ref{fig:lgt2} summarises the    
radiance and velocity evolution in these footpoints,   
labelled A and B in Fig.~\ref{fig:evol} (TRACE image of 14:42:00~UTC).   
A and B are optimised to obtain a high S/N ratio in the smallest region 
still showing relevant coronal velocities. 
The area A consists of three CDS pixels (6$\arcsec\times10\arcsec$), while the   
area B, due to the weak emission of the Fe~{\sc xix} line in the westward   
ribbon, consists of eleven CDS pixels (224 arcsec$^2$). 
\section{Impulsive phase: chromospheric evaporation}\label{impul}    
For both footpoints, the radiances of O~{\sc v} and TRACE 17.1~nm
undergo a sharp increase and a rapid decay   
during the impulsive phase. However,    
Fe~{\sc xvi} and Fe~{\sc xix} present a very different behaviour.  
In A these hotter signatures increase and decrease   
rapidly, while in B Fe~{\sc xvi} becomes   
barely brighter than average and Fe~{\sc xix} begins a slow rise. We   
believe that these differences reflect the different temperature reached by   
the evaporated plasma, being hotter in B 
($>$10$^7$~K, cf.~\S~\ref{grad}). 
\subsection{Motions in the flaring ribbons}   
\begin{figure*}[!ht] 
\sidecaption 
\includegraphics[width=12cm]{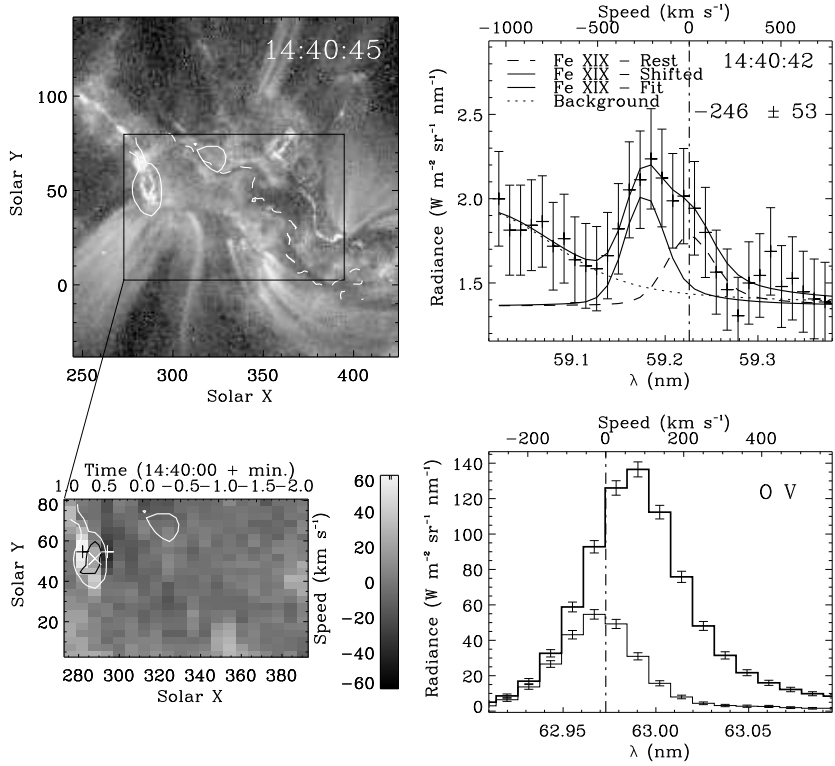} 
   \caption{Eastward ribbon (A) at 14:40:45~UTC. 
   %Early phase of the flare.  
   The top left  panel shows a TRACE image in the 17.1~nm band. 
   The bottom left panel shows a map of the O~{\sc v}   
   LOS velocity within the area encompassed by the black box on the TRACE   
   image. On both panels isocontours of the O~{\sc v}   
   radiances at a level of 5~W~m$^{2}$~sr$^{-1}$ are shown with   
   white thick lines while the thick black isocontours on the velocity 
   map indicate Fe~{\sc xix} radiances at a level of  
   90~mW~m$^{2}$~sr$^{-1}$. 
   The Fe~{\sc xix} profile shown on the top-right panel is obtained by   
   integrating over three pixels along the CDS slit, centred at the 
   location indicated by a white {\bf x} on the velocity map 
	    (area A defined in Fig.~\ref{fig:evol}).   
   %Dashed and solid lines indicate the rest and the shifted components, 
   %respectively. 
   %The dotted line indicates the background profile defined from pre-flare 
   %spectra. The thick solid line indicates 
   %the fitted profile. 
   The single pixel O~{\sc v} profiles on the bottom right panel   
   refer to the locations indicated by a black {\bf +} (thick histogram)   
   and a white {\bf +} (thin histogram) on the 
   velocity map. 
   }   
   \label{fig:start}   
\end{figure*}

{\sc Footpoint A}: Around 14:40~UTC, CDS was scanning the  east flaring 
kernel (A), revealing a strong emission in    
O~{\sc v}. Both Fe~{\sc xvi} and Fe~{\sc xix} clearly increased their emission 
with respect to the background. 
In the 3 pixels defining the kernel A downflows of about 
$20$~km~s$^{-1}$ are measured in O~{\sc v}, but adjacent pixels show 
both strong upflows ($-35$~km~s$^{-1}$, westward of A) and downflows  
(60~km~s$^{-1}$, eastward of A, see Fig.~\ref{fig:start}). The O~{\sc v} 
downflows appear clearly associated with the bright kernel visible in TRACE and 
in the HASTA H$\alpha$ image \citep[see Fig.~1 in][]{Maltagliati-etal:06}. 
The same velocity pattern is also shown by the He~{\sc i} spectra. 
 
At this time the Fe~{\sc xix} spectrum obtained integrating over A shows a blue-shifted 
emission with a 
velocity between $-170$~km~s$^{-1}$ (single component fitting) and  
$-250$~km~s$^{-1}$ (double component fitting, see 
Fig.~\ref{fig:start}, top-right). However, given the noise level in the 
spectrum, it is difficult to establish whether a two component fitting is 
justified. 
% although it seems that a relevant stationary component could be 
%present at this time.  
We thus 
can assume as a lower limit the value obtained with a single component fit. 
The same region also shows upward motions around $-70$~km~s$^{-1}$ in the 
Fe~{\sc xvi} line. Due to the broad NIS-1   
profiles, we have not attempted to separate the strong Fe~{\sc xvi} 
background component, resulting in a likely underestimate of the 
upward velocity. 
 
These upward motions  of hot plasma seem to 
originate in the region between the flow patterns of 
opposite sign observed 
in TR, but from an area where O~{\sc v} motions are still downward 
directed ($\approx20$~km~s$^{-1}$).  
Furthermore, if we consider a semi-circular loop with radial legs, 
the projection effect would amount to about $-2\arcsec$ in the x-direction   
and thus suggest a link  
between the upward motions in Fe~{\sc xix} and 
the even stronger downflows seen in the TR (the target area is on the west hemisphere,  
the loop direction is east-west, and we assume the Fe~{\sc xix} 
emission from a region $\approx$5\,000 km above that of O~{\sc v} and He~{\sc i} ).  
%A link between the upflows in 
%Fe~{\sc xix} and those seen in the TR lines would require the loops to be 
%strongly slanted westward. 
%===================================================================   
%========================================================================   
 
\begin{figure*}[!t] 
\sidecaption 
\includegraphics[width=12cm]{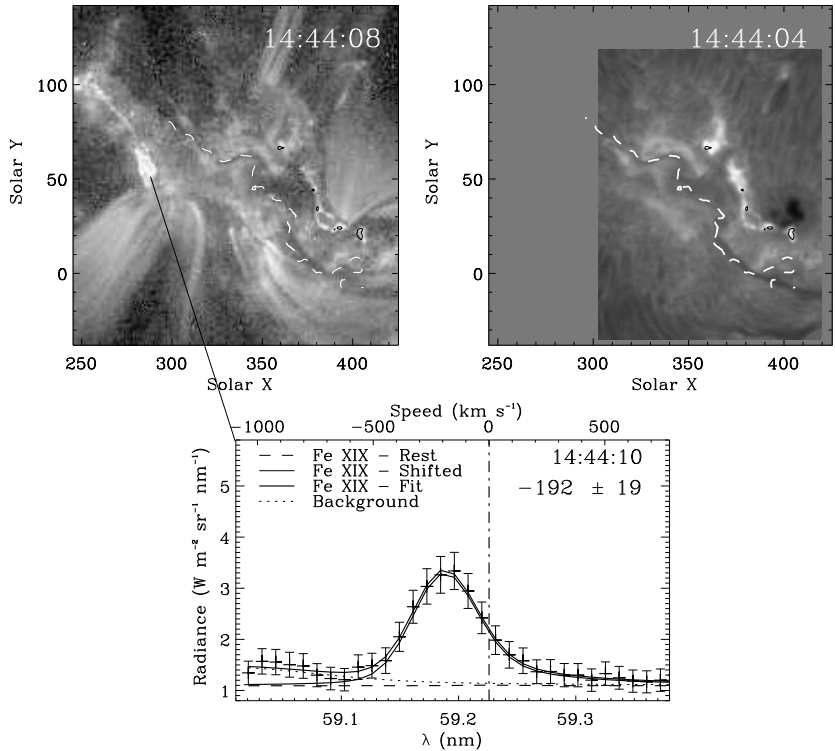}   
   \caption{ 
   %Chromospheric evaporation in the  
   Eastward ribbon (A) at 14:44~UTC. The top left   
   panel shows a TRACE image in the 17.1~nm band while the    
   H$\alpha$ image closest in time is shown on the   
   top right panel. 
   On both images the thin black contours indicate chromospheric   
   downflows of 4~km~s$^{-1}$.  
   The Fe~{\sc xix} line profile shown on the bottom panel is obtained by   
	    integrating over three pixels along the CDS slit   
	    (area A defined in Fig.~\ref{fig:evol}).   
	    }   
   \label{fig:impvel2}   
\end{figure*}

Around 14:44~UTC,   
CDS scanned again this flaring kernel (Fig.~\ref{fig:impvel2}).  
%As mentioned earlier, due to the    
%stronger emission of the Fe~{\sc xix} line at this time and position,    
%we could average over only three CDS pixels to get a significant S/N value.    
The resulting Fe~{\sc xix} profile (bottom panel of the figure) shows   
a strong blue-shifted component dominating   
the whole profile. The corresponding  velocity is of about $-$200~km~s$^{-1}$.  
Upflows of $-70$~km~s$^{-1}$  are still measured in the coronal Fe~{\sc xvi}. 
The same velocity pattern seen in O~{\sc~v} and He~{\sc i} during the 
previous passage is still discernible, with the downflows now  
weaker ($\approx20$~km~s$^{-1}$) and unchanged upflows.  
 
Finally, strong coronal upflows are still visible   
in the kernel A when CDS scans the area a third time during the   
impulsive phase, around 14:47:30~UTC ($-170$~km~s$^{-1}$ with a double component 
fitting and $-100$~km~s$^{-1}$ with a single component fitting). 
O~{\sc~v} and He~{\sc i} do not show any downflow anymore, while the nearby  
upflow is still present.\\ 
 
{\sc Footpoint B}: This footpoint is within the DST FOV.  
Starting from around 14:41~UTC,  
the difference of H$\alpha$ images   
   obtained at both $\Delta\lambda$=$\pm~60$~pm and $\pm~150$~pm    
   shows a red-wing emission excess in   
correspondence of the flaring kernels, 
resulting in a downflow velocity of 4 to 10~km~s$^{-1}$ 
(cf.~\S~\ref{speed}). 
 
\begin{figure*}[!ht] 
\sidecaption 
\includegraphics[width=12cm]{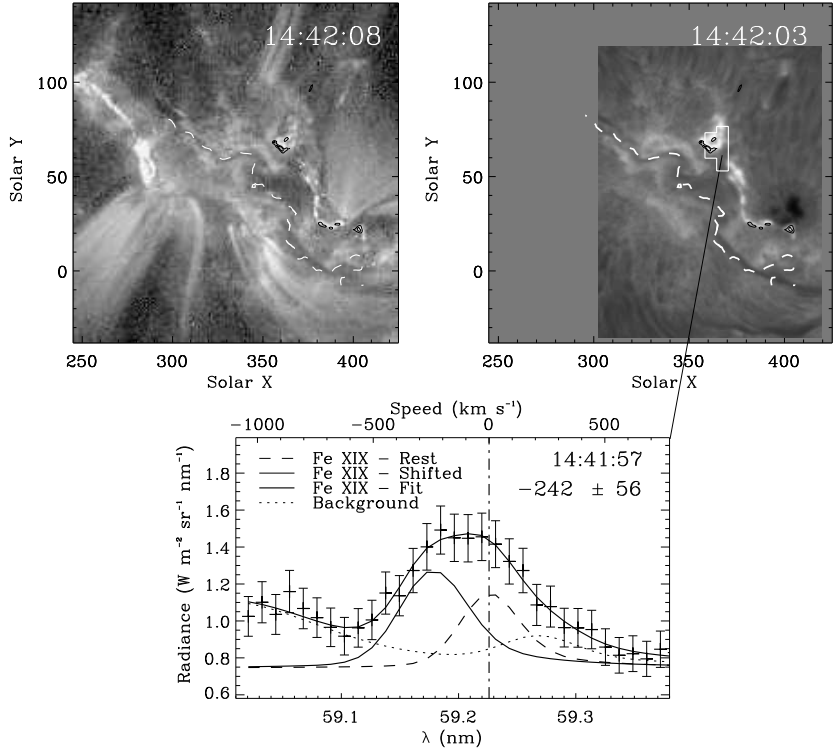} 
   \caption{ 
   %Chromospheric evaporation in the  
   Westward ribbon (B) at 14:42~UTC. Panels are as in Fig.~\ref{fig:impvel2}.    
   On both images the thin black contours indicate chromospheric   
   downflows of 4~km~s$^{-1}$ and 7~km~s$^{-1}$.   
   The Fe~{\sc xix} line profile shown on the bottom panel is obtained by   
   integrating over the area within white solid lines in the H$\alpha$   
   image (area B defined in Fig.~\ref{fig:evol}).  
   A single component fit (plus background) would result in   
   an upflow speed of ($-125\pm50$)~km~s$^{-1}$   
   }   
   \label{fig:impvel}   
\end{figure*}

Around 14:42~UTC this area was rastered over by CDS. 
Velocities  in O~{\sc v} and He~{\sc i} indicate only downwards motions 
($\approx$20~km~s$^{-1}$) outlining the tiny bright features visible at 
the higher spatial resolution afforded by H$\alpha$ data.  
Chromospheric downflows are measured at the   
same time in the same area, and are indicated with   
thin black contours in the upper panels of Fig. ~\ref{fig:impvel}.   
Upflows of more   
than 200~km~s$^{-1}$ were measured in Fe~{\sc xix} by averaging over  
the whole footpoint area. The   
corresponding profile of the Fe~{\sc xix} line is shown 
in the bottom panel of the same figure. The fitting procedure   
reveals a blue-shifted component of the same strength   
of the stationary component, with a speed of $-$240~km~s$^{-1}$.  
A single component fit provides a velocity of about   
$-$125~km~s$^{-1}$. 
 
During the next raster at  14:45:23~UTC, CDS    
still measures strong upward speeds in the Fe~{\sc xix} line 
(Fig.~\ref{fig:lgt2}),   
cospatial and cotemporal with downflows in the low chromosphere (H$\alpha$).   
The radiance of the blue-shifted component is comparable    
to that of the stationary component also at this time.  
%Hence in this footpoint, the   
%chromospheric evaporation (characterised by a coronal upflow and a   
%simultaneous and co-spatial chromospheric downflow) persists at least for  
%200~s, i.e., a time comparable with the FWHM    
%of the HXR proxy shown in Fig.~\ref{fig:sxr}. 
%This is the first time that the  oppositely directed 
%motions measured in low-chromosphere and in hot flare lines, are 
%verified to be co-spatial and co-temporal.\\ 
At later times, neither the Fe~{\sc xix} nor the H$\alpha$ data 
show any relevant motions. Hence in this footpoint, the   
oppositely directed 
motions measured in low-chromosphere and in hot flare lines, are 
verified to be co-spatial and co-temporal and to persist for at least 
200~s.\\  
 
%===================================================================   
%===================================================================   
Contrary to what could be expected, the Fe~{\sc xvi} line does not   
show any upward velocity within the area B. However, we note that  
at the times when strong   
motions are measured in Fe~{\sc xix}, the radiance of the Fe~{\sc xvi} is   
practically unchanged with respect to its pre-flare values. This also suggests  
that the evaporated plasma is too hot to generate a shifted component strong   
enough to be measurable with respect to the  
Fe~{\sc xvi} background emission. 
 
Comparing the two footpoints, we notice that the \ion{Fe}{xix} blue-shifted 
component is clearly dominating  over the   
stationary one only in the eastward ribbon (A), possibly due  
to the smaller area used to obtain reliable line profiles.    
This fact may then support the assumption that    
multiple thin loops are present in the flaring ribbons at any time, and   
that they are heated successively,    
as in the simulations of \cite{Warren:05}. This is further   
confirmed by TRACE  and H$\alpha$ images that show very clearly    
the existence of multiple footpoints within the flaring ribbons,   
brightening at different times.   
Note also that during the gradual phase (see \S~\ref{grad})    
post-flare loops of 1$\arcsec$~to~2$\arcsec$ diameter clearly emerge as 
independent features in the TRACE images.   
 
%A special mention should go to the motions of the plasma at TR temperatures as 
%obtained from the O~{\sc v} and He~{\sc i} spectra.  
Finally, for both footpoints, the large upflows measured in the Fe~{\sc xix} line are 
spatially related to patches of downflows in the TR, themselves closely 
associated to the flaring kernels. However, a significant upflow pattern lasting 
at least 600~s is also present nearby footpoint A, but we are at lack 
of an obvious explanation for it.  
%Models by \citet{Fisher-etal:85a} result in upflows or 
%downflows, on both TR and chromosphere, depending on the energy flux of the 
%impinging electrons, with downflows predicted for large energy fluxes  
%(explosive evaporation). 
%   
\subsection{Momentum balance}\label{moment}   
     
Hydrodynamical simulations of explosive chromospheric evaporation predict 
the equality of momenta between the hot plasma moving upward and  the 
dense cold plasma moving downward  
during the impulsive phase of a flare \citep{Fisher-etal:85b}. If the   
oppositely directed flows that we measure during the impulsive phase   
indeed signal chromospheric evaporation, we should be able to verify such   
equality. Earlier attempts have been performed, using BCS/SMM coronal   
data, by \citet{Zarro-etal:88} and \citet{Canfield-etal:90a}. These authors find  
agreement between momenta of the oppositely directed plasma   
(integrated over the whole impulsive phase) within one   
order of magnitude, although the lack of spatial resolution in coronal data   
leaves open the issue of co-spatiality of the flows.    
We have shown that chromospheric downflows and 
coronal upflows are co-spatial and co-temporal at least in the 
case of footpoint B, so we can compare the ``instantaneous'' momenta at 
14:42~UTC in this footpoint.  
%For the plasma evaporated upward into the 
%corona we use the values derived from Fe~{\sc xix} spectra. 
 
The calculation of the downward chromospheric ($P_{down}$) and the upward coronal  
($P_{up}$) momenta is described in Appendix~\ref{app},  
where the uncertainties of the measured or 
assumed physical parameters are taken into account. We also note that  
the calculations of both momenta involve totally 
independent sets of physical quantities. For the downward 
momentum we obtain  
$5.1~\times~10^{19}$~g~cm~s$^{-1}~\le~P_{down}~\le~4.6~\times~10^{20}$~g~cm~s$^{-1} $. 
%The lower and upper limits refer to the minimum 
%and maximum values of the {\it pre-flare} hydrogen density, the  
%estimated chromospheric velocity 
%and the condensation thickness.  
The range of values depends from the {\it pre-flare} hydrogen 
density, the chromospheric velocity and the condensation thickness 
that we can estimate only within a factor of two. 
For the upward momentum, we obtain 
$5.9~\times~10^{19}$~g~cm~s$^{-1}~\le~P_{up}~\le~4.9~\times10^{20}$~g~cm~s$^{-1}$, 
where the lower and upper limits refer to 
the velocity and density values obtained assuming a single or a 
double-component fit of the Fe~{\sc xix} line, an iron abundance varying    
between photospheric and coronal values and the minimum and maximum 
value of the coronal filling factor. Possible departures from ionisation 
equilibrium can also contribute to the uncertainty of the result. 
In principle we could also attempt to check the momentum balance at 14:45:23~UTC,   
when we still  measure co-spatial coronal and chromospheric flows on   
the area analysed above.   
However, at this time the chromospheric density   
has already increased to flare values, difficult to estimate without an   
{\it ad hoc} flare model, which we cannot obtain for lack of suitable   
spectral signatures.  
%===================================================================   
%==================================================================   
 
The equality within one order of magnitude of  
the ``instantaneous'' chromospheric and coronal momenta 
supports the explosive chromospheric evaporation model, even though the flare analyzed is  
quite small and thus unlikely to provide the 
large energy flux requested by the simulations.  
%It can be seen that the    
%``instantaneous'' momenta of the coronal and chromospheric moving plasma   
%are equal within one order of magnitude.  
%It should be noted that the calculations of both momenta involve totally 
%independent sets of physical quantities. 
% 
%Thus,   
%our observations support the chromospheric evaporation   
%models during the impulsive phase of this flare. In fact, for a time when we   
%measure chromospheric and coronal flows,    
%we show not only that they originate from the same area, but also that their   
%momenta are equal within one order of magnitude.   
More in general, we note that even with simultaneous, spatially resolved 
observations we cannot obtain a verification of the momentum balance  
better than one order of magnitude  
due to inherent uncertainties in the physical parameters. 
\section{Gradual phase}\label{grad}    
The gradual phase of the flare starts around 14:48~UTC.  As clearly   
visible in Fig.~\ref{fig:lgt2}, during this phase no significant upflows   
in coronal lines are measured (nor corresponding chromospheric   
downflows). This indicates the end of chromospheric evaporation   
within the flare.   
 
\begin{figure*}[!ht] 
\sidecaption 
\includegraphics[width=12cm]{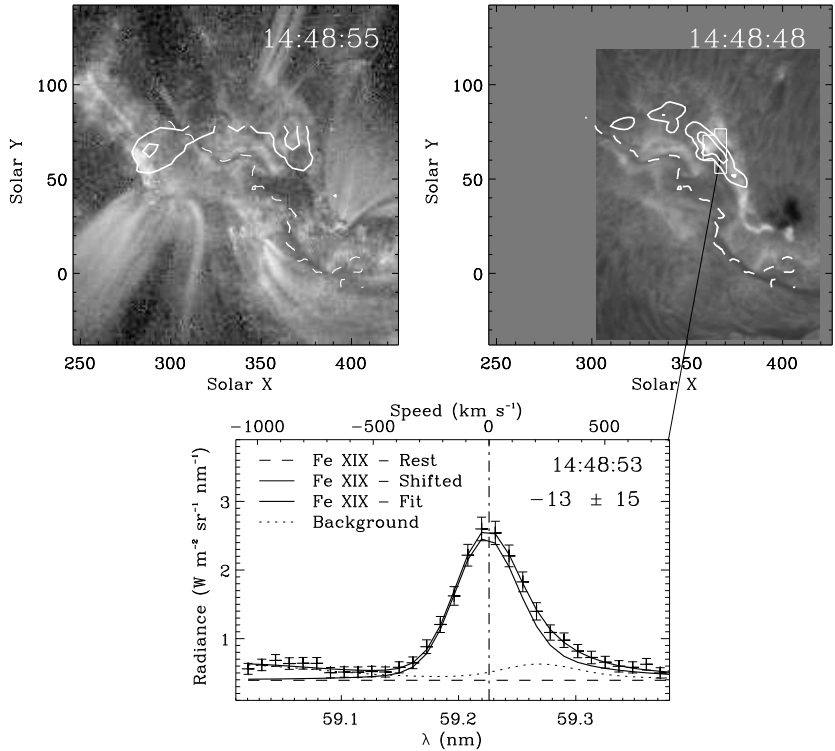}   
   \caption{Gradual phase at a time when the CDS slit was crossing the   
   western   
            footpoint. The top left panel show a TRACE image in the 
	    17.1 nm band while the H$\alpha$ image closest in time 
	    is shown on the top right panel.     
	    On the TRACE image isocontours of the Fe~{\sc xix} radiance at   
	    levels of (0.11, 0.22)~W~m$^{-2}$~sr$^{-1}$ are indicated by   
	    thick white solid lines.    
	    On the H$\alpha$ image, instead, the white thick contours refer to   
	    a RHESSI image obtained between 14:50 and 14:51~UTC in the   
	    6~to~12~keV band.    
	    The contours are traced at levels of 40~\% and 75~\% of the maximum. 
	    The Fe~{\sc xix} line profile shown on the bottom panel   
	    is obtained by integrating over the areas within white solid thin   
	    lines in the H$\alpha$ image (area B).   
	    In this case the radiance of the rest component (dashed line)   
	    was set to zero. The dotted line indicate the background (lines   
	    plus constant background) defined from pre-flare spectra. The thick 
	    solid line indicate the fitted profile.    
	    }   
   \label{fig:grad}   
\end{figure*}

RHESSI data were available from 14:50~UTC. Images were   
reconstructed in the 3~to~12~keV band, at 14:50, 14:52 and 15:13~UTC, 
with an integration 
time of 60~s using the {\it Pixon} algorithm of the standard RHESSI software 
\citep[]{Maltagliati-etal:06}. In  
Fig.~\ref{fig:grad} we show the TRACE and H$\alpha$ images  at 14:48~UTC, 
with overplotted, respectively, the contours of Fe~{\sc xix} radiance and      
of the RHESSI image in the 6~to~12~keV band at 14:50~UTC 
(thick solid white line). The RHESSI 
emission encompasses the western ribbon and extends eastward outlining a    
loop structure that is also partially visible from the Fe~{\sc xix}   
radiance isocontours   
(note that the loop top lies outside the CDS FOV), but   
no emission is registered on the eastern footpoint (A). Thus, also RHESSI data   
outline a different behaviour of the two flaring ribbons.   
   
\begin{figure*}[!t] 
\resizebox{\hsize}{!}{\includegraphics{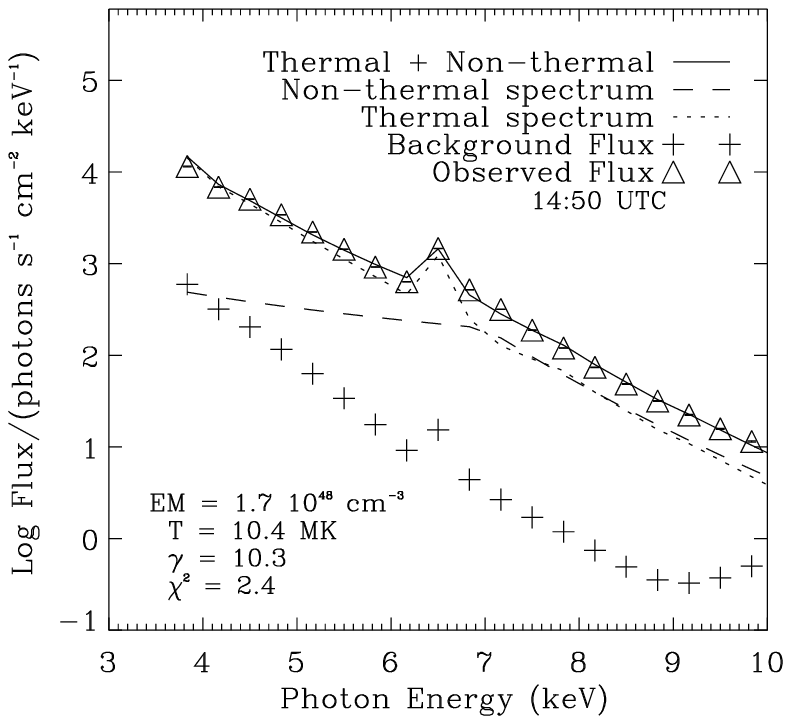}\includegraphics{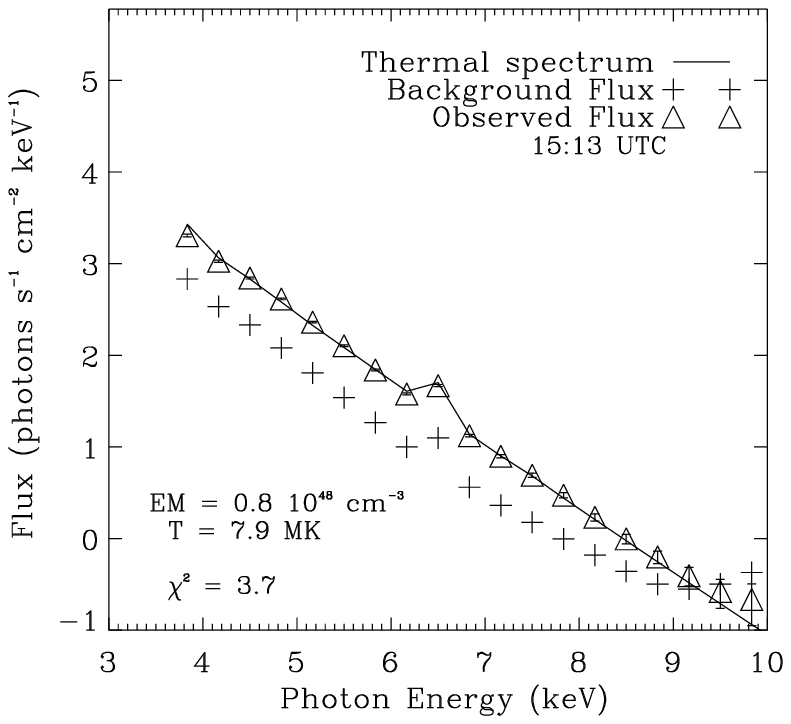}} 
   \caption{RHESSI spectra in the 4 to 10~keV energy range at 14:50~UTC   
   (left panel) and at 15:13~UTC (right panel). Notice the presence of a   
   broadened emission feature at about 6.6~keV corresponding to a group of   
   emission lines due to Fe~{\sc xxiv~--~xxvi}.}   
   \label{fig:rhessi}   
\end{figure*}

The corresponding   
integrated RHESSI spectrum mainly describes    
the emission of an area closely overlapping kernel B.   
This spectrum (Fig.~\ref{fig:rhessi},    
left panel) reveals the broadened emission line feature   
around 6.6~keV, due to lines of   
highly ionised iron (Fe~{\sc xxiv~--~xxvi})   
observable in RHESSI data at temperatures above $\approx   
10^7$~K \citep{Phillips:04}.   
Indeed, the best fit to the spectrum indicates a thermal   
component of $T\approx 10^7$~K,    
as well as a non-thermal one with a spectral index   
$\gamma \approx10$.    
$\gamma$ values of the same order have already been   
observed in RHESSI data during the decay    
phases of small flares \citep{Krucker-etal:02, Hannah-etal:04}. The   
presence of non-thermal electrons at these low energies, however, is 
not obviously linked to any evaporation process, as no upflows are measured   
in the coronal signatures (bottom panel of Fig.~\ref{fig:grad}).   
Around 23 min. later (Fig.~\ref{fig:rhessi}, right panel)   
the spectrum exhibits only a thermal component with a temperature   
$\approx 8\times10^6$~K.    
%===================================================================   
%===================================================================   
   
A coronal plasma heated above $10^7$~K in kernel B might explain the 
very different behaviour displayed by the two footpoints, both in radiance 
and velocity. In A, after 
the end of the impulsive phase, both coronal signatures undergo a rapid 
decay, 
%consistently with the    
%expectation of maximum emission at the end of the evaporation process   
%\citep[]{Hudson-Ryan:95,Brosius:03}.   
%On the contrary,   
while in B the radiance of Fe~{\sc xix} continues its increase up to 
15:05~UTC, i.e., well into the gradual phase. The Fe~{\sc xvi} shows a 
similar trend,  but with a more significant delay, reaching its maximum 
radiance in B almost one hour after the GOES peak. This seems consistent with  
the rapid cooling of ``hot'' plasma ($>10^7$~K), evaporated during the 
impulsive phase, to Fe~{\sc 
xix} temperatures, and successively, to values characteristic  of Fe~{\sc xvi}  
emission.  
 
In this same footpoint, downward directed motions are measured in both  
Fe~{\sc xix} and Fe~{\sc xvi} around the time of their peak emission, although
the amplitudes are close to the error level (for Fe~{\sc xix}, 
(20~to~70~$\pm~50$)~km~s$^{-1}$ for a double-component fitting  
and (6~$\pm$~12)~km~s$^{-1}$ for a single-component fit; for Fe~{\sc xvi}, 
(20~$\pm~12$)~km~s$^{-1}$). Downflows of 
$\approx(20\pm$~9)~km~s$^{-1}$, are further observed in  O~{\sc v} to outline
the B footpoint around the time of maximum O~{\sc v} emission during the
gradual phase ($\approx$15:42 UTC). No obvious downflows are instead measured
in A during the gradual phase.  
 
%The downflows measured in B   
%can be interpreted as cooling-associated flows, much as in the case reported  
%by \citet[]{Czaykowska-etal:99}, based on similar CDS observations.    
%However, while relevant   
%downflows during the cooling of postflare loops are present in several   
%hydrodynamical models \citep[]{Doschek-etal:82, Tsiklauri-etal:04,   
%Bradshaw-etal:04}, they usually appear only when the plasma has cooled down to 
%temperatures around (1~to~2)$\times10^6$~K. The motions measured in the hotter  
%Fe~{\sc xix} would then require a different explanation. 
%The downflows measured in both Fe~{\sc xvi} and Fe~{\sc xix}   
%can be interpreted as cooling-associated flows, much as in the case reported  
%by \citet[]{Czaykowska-etal:99}, based on similar CDS observations.    
%However, while relevant   
%downflows during the cooling of postflare loops are present in several   
%hydrodynamical models \citep[]{Doschek-etal:82, Tsiklauri-etal:04,   
%Bradshaw-etal:04}, they usually appear only when the plasma has cooled down to 
%temperatures around (1~to~2)$\times10^6$~K. Maybe the flows   
%we observed in Fe~{\sc xix} can be compared with the    
%downward motions of   
%flaring plasma around 10$\times10^6$~K reported by \citet[]{McKenzie:00} for 
%the case   
%of bright features in limb SXT data. He suggests that such motions evidence   
%magnetic field line shrinkage (i.e., reconnection outflow) in flaring   
%supra-arcades.    
 
Finally, combining gradual phase    
RHESSI data with CDS measurements up to 14:50~UTC, we evaluate 
the contribution of the evaporated chromospheric plasma in filling the flare   
loops. Using the Fe~{\sc xix} data, we can estimate the total number of 
electrons transported with evaporation during the impulsive phase,  
assuming a symmetric evaporation on the two footpoints:  
   
\begin{equation}   
N_{up} =  n_{\rm e}~2~S_{cor}~f~v_{cor}~\Delta t,   
\label{eq:massa_up}   
\end{equation}   
   
\noindent where $n_{\rm e}$ is the electron density  of the upflowing plasma,   
$f$ the coronal filling factor, 
$2~S_{cor}$ the area of the observed upflows (both footpoints), $v_{cor}$ 
the velocity, and $\Delta t$ the duration of the evaporation. The electron 
density and the   
upward velocity are the same as in \S~\ref{moment} (see also Eq.~\ref{eq:ne}) 
and depend on whether we   
use one ($v_{cor}=-125$~km~s$^{-1}$) or two ($v_{cor}=-240$~km~s$^{-1}$)   
components to fit the Fe~{\sc xix} line profile at 14:42:03~UTC and upon the 
assumed iron abundance. 
We assume such densities and velocities as typical throughout the   
evaporation.   
$2~S_{cor}=2\times1.2\times~10^{18}$~cm$^2$ accounts for  the   
fact that the plasma is flowing from both footpoints.    
Finally, $\Delta t = 300$~s is taken equal to the duration of chromospheric   
downflows, for which we have a better temporal resolution.   
   
With these values, for the two cases of single and double fitting and for both 
iron abundances of 7.5 and 8.1, we   
obtain the minimum and maximum estimate of    
$N_{up}=$~[3.8,10.]$\times10^{37}\sqrt{f}$~electrons.   
   
The above values can be compared with the total number of electrons in the   
entire loop system after the end of the evaporation phase.   
This can be written as:   
\begin{equation}   
N_{loop} = \sqrt{\frac{EM_{RHESSI}}{f~S_{cor}~L_{loop}}} f~S_{cor}~L_{loop},   
\label{eq:massa_loop}   
\end{equation}   
\noindent where $EM_{RHESSI}=0.8\times10^{48}$~cm$^{-3}$ is the emission   
measure obtained from the RHESSI spectrum at 15:13~UTC (only thermal component,   
see Fig.~\ref{fig:rhessi}). From the TRACE images around 16:20~UTC    
a loop length $L_{loop}$ of $\approx9.2\times10^{9}$~cm can be derived,   
from which $N_{loop}=9.4\times10^{37}\sqrt{f}$~electrons. Since both $N_{up}$   
and $N_{loop}$ have the same dependence on $\sqrt{f}$,    
the evaporated material can account for  
40~\% up to 100~\% of the material inside the loop system.  
This implies that the evaporated   
plasma largely contribute to the density of the post-flare loops.   
 
%===================================================================   
%===================================================================   
%     
\section{Cooling of post-flare loops}\label{cool}   
\begin{figure}[!t] 
\resizebox{\hsize}{!}{\includegraphics{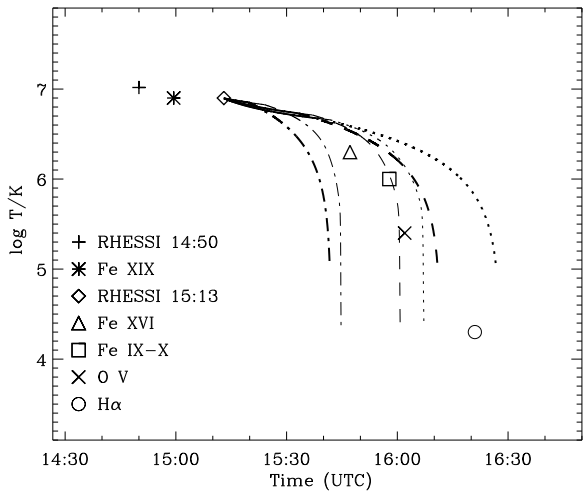}} 
   \caption{Cooling curve obtained from the time of maximum appearance of   
   progressively colder features. RHESSI temperature at 14:50~UTC   
            refers to the thermal component (see left panel of   
	    Fig.~\ref{fig:rhessi}). Each set of curves (thin and thick)   
	    refer to three different values of the volumetric filling factor   
	    $f$ (0.1, dot-dashed; 0.3, dashed; 0.5, dotted). The solid part   
	    of the curves indicates the times when conduction dominates,   
	    while the non-solid part indicates the time when the plasma cools   
	    radiatively. 
	    The thick curves are obtained assuming   
	    constant density only during the conductive phase, while the thin   
	    curves are obtained assuming constant density through the entire   
	    cooling (see text).}   
   \label{fig:cool}   
\end{figure}

The whole evolution of the flare outlines the filling of magnetic loops by   
plasma evaporated from the chromosphere during the impulsive phase. Such   
loops become visible in all of our signatures during the extended gradual   
phase, with progressively cooler features appearing at later times (see   
Fig. \ref{fig:evol}).   
Fig.~\ref{fig:cool} reports the typical temperatures corresponding to our   
signatures, assumed as those corresponding to the maximum of the   
contribution function, vs. the peak time of the loop radiances.    
Moreover,   
the temperature of the thermal spectrum used to fit RHESSI data is also   
reported at the two times of acquisition.   
   
We compare here the temporal evolution defined by the data   
with the results of the analytical formulae provided by \citet{Cargill-etal:95}   
for the cooling time of a flare plasma. The first RHESSI spectrum at 14:50~UTC   
still shows a non-thermal component (see \S~\ref{grad}), suggesting further   
energy input at the beginning of the gradual phase. Hence we begin   
the analysis of the cooling at 15:13~UTC, when RHESSI shows only a thermal   
component. The first data point corresponds to a $T\approx 8\times10^6$~K, and an   
average electron density of $8.5\times10^9$~cm$^{-3}$ (for $f=1$).   
We used the formulae provided for the static case since  
after 15:13~UTC no evaporative motions are visible anymore.   
The calculations were performed following two different approaches.  
 
In the   
first case a constant density was assumed throughout the entire cooling period 
and   
the complete radiative loss function of \citet{Rosner-etal:78} was used in the   
calculations. For this purpose Eqs.~7S and 8S of \citet{Cargill-etal:95} have   
been generalised to any value of the exponent $\alpha$ of the power law.  
Using such equations, we   
computed  the time until when conduction dominates over radiation. The   
temperature evolution of the cooling plasma was obtained through Eq.~3a 
of \citet{Cargill-etal:95} before this time and Eq.~4 after,     
with the appropriate   
values for each piece of the radiative loss function of \citet{Rosner-etal:78}.   
The thin lines in Fig.~\ref{fig:cool} show the results of these calculations   
for three   
values of the filling factor $f$ (0.1, dot-dashed; 0.3, dashed;    
0.5, dotted). The solid part of the curves   
indicates the times when conduction dominates, while the non-solid part indicates   
the time when the plasma cools radiatively.   
   
%Because the constant density assumption may be questionable,  
In the second   
approach we assumed that the density is constant only during the conductive   
phase while $n \propto \sqrt{T}$ in the radiative phase \citep{Serio-etal:91}.   
We also assumed the simplified radiative loss function   
$P_{rad}=1.7\times10^{-19}n^{2}T^{-1/2}$ that fits, within a factor two, the   
CHIANTI radiative loss function for coronal abundances for    
$5~\le~\log(T/{\rm K})~\le~7$. During the time   
dominated by conduction, the calculations are similar to the previous case.   
After this time, being $\alpha=-0.5$, we can use Eq.~6 of \citet{Cargill-etal:95} 
to calculate the temperature evolution during the time when radiative cooling   
dominates. The results are represented by the thick curves in   
Fig.~\ref{fig:cool}, for the same values of $f$.   
Again, the solid part represent the conductive cooling while the non-solid part is   
the radiative one.    
   
Both approaches show that the observed  
curve can be explained by the cooling of the hot flare plasma injected into the   
loop system only if a volumetric filling factor of $\approx$~0.2~to~0.5 is   
assumed, a value well in agreement with the estimated   
filling factors of cooling loops given by \citet{Aschwanden-etal:03}.    
We notice that the estimated cooling time in the temperature range 
$4~\le~\log(T/{\rm K})~\le~5$ is much shorter than   
the observed one. This is expected because our initial assumptions are   
no longer valid in this range, in particular due to recombination of hydrogen.   
 
\section{Conclusions}\label{concl}   
We presented comprehensive observations of a small eruptive flare (GOES   
class C2.3) that developed on 2002 August 11 in NOAA region 10061. A wide 
array of instruments and diagnostics allowed us to follow the evolution of the flare from the   
chromosphere to the corona during both the impulsive phase and 
the long gradual phase. The eruption of a portion of a filament,   
located on the apparent magnetic inversion line, was signalled by an upward velocity   
of about $ - 30$~km~s$^{-1}$ detected at  14:36:33~UTC by CDS    
in the O~{\sc v} line.   
A few minutes later the footpoints of a large loop system, subsequently visible    
also in the TRACE images, began to brighten in all    
signatures, reaching their   
maximum brightness around 14:42~UTC.    
%===================================================================   
%===================================================================   
 
During the impulsive phase of the   
flare, large coronal upflows of $-200$~km~s$^{-1}$ or more are measured from the   
Doppler shift of the Fe~{\sc xix} line profiles on the ribbons at the footpoints   
of the flaring loop system.    
Interestingly, the two footpoints show  quite a different behaviour. Footpoint A 
shows a fast (within our time resolution) rise of the radiances of all our 
signatures, 
followed by a slower decay of the Fe~{\sc xvi} and Fe~{\sc xix} emission. 
On this footpoint, during the impulsive phase, we can select a small area (3 
CDS pixels) where a blue-shifted component clearly dominates the Fe~{\sc 
xix} profile. Upward motions of about $-70$~km~s$^{-1}$ (likely
underestimated) are also measured from the Fe~{\sc xvi} CDS data. 
%the Fe~{\sc xix} profiles show a blue-shifted    
%component clearly dominating over the stationary one and  
These coronal upward motions occur in the location where downflows are observed 
in the O~{\sc v} and He~{\sc i} lines for at least 200~s 
%These motions, both in Fe~{\sc xix} and in TR lines,  
%last for at least 200~s. 
but, unfortunately, no H$\alpha$ Dopplergrams are available for this 
footpoint. 
 
On footpoint B instead, the radiances of Fe~{\sc xix} and 
Fe~{\sc xvi} lines 
increase very slowly and reach their maxima respectively 20~min and 60~min 
after the start of the flare. RHESSI data at 14:50~UTC show the emission in the 
3 to 12~keV band to be mainly associated to this footpoint, with the 
spectrum 
indicating a temperature of the thermal component of 10.4~MK, i.e. larger  
than the Fe~{\sc xix} formation temperature (8~MK). This would explain the 
slow rise of Fe~{\sc xix} and Fe~{\sc xvi} radiances as due to progressive 
cooling of this ``hot'' plasma.  
During the impulsive phase, the blue-shifted and the stationary component 
of the Fe~{\sc xix} line profile on this area are of comparable strength, 
and no clear motion is measured from  the Fe~{\sc xvi} line.  
%This    
%suggests the presence of multiple   
%thin loops heated successively as in the simulations of \cite{Warren:05}.   
The O~{\sc v} and He~{\sc i} lines show downward  
($\approx20$~km~s$^{-1}$) motions outlining the 
flaring kernels. 
For this footpoint, within the FOV of the UBF imager, we   
observed simultaneous H$\alpha$ downflows that we estimated to be about    
($7\pm3$)~km~s$^{-1}$. To our knowledge, this is the first time that co-spatial,  
co-temporal, oppositely directed flows during the impulsive phase of a flare   
are measured in low chromosphere and corona, supporting the model of explosive  
chromospheric evaporation. 
 
Such direct measurement gave us the possibility to compare the   
``instantaneous'' momenta of the coronal and chromospheric moving plasma,   
whose equality is required in the hypothesis of explosive 
evaporation. We found that such momenta   
are equal within one order of magnitude.  
Previous works \citep{Zarro-etal:88, Canfield-etal:90a} verified the equality 
of the momenta within the same limits using data without spatial resolution and 
integrating over the whole impulsive phase. 
Although we can utilise spatially resolved data, it remains very difficult to obtain 
more accurate estimates of the momentum balance, due to the inherent 
uncertainties in  physical parameters 
such as coronal electron density and abundances, chromospheric hydrogen 
density and condensation height. To improve the knowledge of these parameters 
a larger number of spectral lines both in the EUV (including density 
diagnostics) and in visible/near-infrared (to build a semi-empirical model of 
the chromosphere) are necessary. However, this is likely to jeopardise the 
temporal and spatial resolution needed to study the impulsive phase of a flare. 
New space instruments with larger effective areas, such as the forthcoming 
EIS on Solar B, should improve the situation on the EUV side. 
   
A further element in support of the chromospheric evaporation was given by   
RHESSI data acquired during the gradual phase, from which we obtained   
an estimate of the mass in the coronal loop system. Combining the velocity   
and radiance data from CDS, we could also estimate the mass evaporated during   
the impulsive phase of the flare, and found that it provided  
a large contribution to the density of the X-ray emitting coronal 
material. 
%The presence of the strong downflows (20~to~70~km~s$^{-1}$) measured in the 
%Fe~{\sc  xix} line during the early gradual phase is a prominent feature of 
%this flare. However, their interpretation remains open, since cooling-associated 
%flows predicted by hydrodynamical models usually appear only at lower   
%temperatures.   
   
Finally, during the extended gradual phase, the magnetic loops filled by   
evaporated plasma become visible in all of our   
spectral lines and band-passes, with progressively cooler features appearing 
at later times.       
The observed temporal   
evolution can be reproduced using  the analytical formulae of   
\citet{Cargill-etal:95}. Initially the cooling is   
dominated by conduction, while radiative losses take over later on.   
A cooling time comparable to the observed one can be obtained    
only assuming a volumetric filling factor of $\approx$~0.2~to~0.5.   
This in turn is consistent with the the presence of multiple thin loops heated 
successively, suggested by the observed dominant blue-shifted 
component in the \ion{Fe}{xix} line profile only when we can use a small area  
to obtain reliable data. 
%suggests the presence of multiple   
%thin loops heated successively as in the simulations of \cite{Warren:05}.   
   
To end this paper, we would like to make some remarks about the complexity of the 
TR behaviour. While our observations show downflows in the flaring kernels also from TR lines, 
in qualitative agreement with  some hydrodynamical 
simulations of explosive evaporation \citep{Fisher-etal:85a,Fisher-etal:85b}, we note that 
recent hydrodynamic models of chromospheric evaporation  
\citep{allred-etal:05} always predict {\it upflows} for the TR plasma.  
Several observations    
obtained with CDS during the impulsive phase of flares, with programs   
privilegeing either the spatial or the temporal resolution, offer a wide range of 
results. Upflows have been reported by    
\citet{Teriaca-etal:03}, using an observational sequence similar to the one   
presented in this paper, and \citet{Brosius:03}, that adopted a very high   
cadence program, but with a narrow field of view and limited spatial resolution.  
Upflow is also 
measured in our data during the impulsive phase, in the eastern footpoint 
nearby the area A but not appear to be
obviously connected to the coronal upflows.. Downflows are   
instead reported by   
\citet{Brosius:04}, using the same observing program as   
\citet{Brosius:03}, and \citet{Kamio-etal:05}   
that analyse four flares in a larger FOV but with intermediate temporal   
resolution. Although at much lower spatial and temporal   
resolution, the same contradictory indications    
were obtained with SMM and OSO-8 instruments   
\citep[see the discussion in ][]{Teriaca-etal:03}. 
It is likely that the behaviour of the TR is strongly related to the 
structure of the atmosphere depending upon the local magnetic topology. Space 
observations with substantially higher spatial resolution  than now ($\le1''$) 
are most 
likely necessary to better understanding the dynamics of the TR and provide 
constraints to theoretical models. Further, we would like to point out  that 
hydrodynamical simulations are generally computed for energetic flares 
(GOES class X-M) with a much harder spectrum for the impinging electrons 
($\gamma$ from 4 to 6) than normally observed with RHESSI ($\gamma$ from 
7 to 10) in small flares such as the one studied here.

%%%%%%%%%%%%%%%%%%%%%%%%%%%%%%%%%%%%%%%%%%%%%%%%%%%%%%%%%%%%%%%%%%%%%%%%%%%%%%%   
%%%%%%|   
%%%%%%|  ACKNOWLEDGEMENTS   
%%%%%%|   
%%%%%%%%%%%%%%%%%%%%%%%%%%%%%%%%%%%%%%%%%%%%%%%%%%%%%%%%%%%%%%%%%%%%%%%%%%%%%%%   
   
\begin{acknowledgements}   
The authors thank Prof. E. Priest for suggesting calculating the amount   
of plasma evaporated into the loop and Dr. K. Wilhelm 
for carefully reading the manuscript. We are also grateful to Dr. V. Andretta 
and Dr. U. Sch\"{u}hle for useful comments and suggestions. 
Many thanks are due to the NSO and CDS staffs for their help and assistance  
in acquiring the data here analysed. We would also like to    
thank the RHESSI staff for their help in analysing the data.   
SOHO is a mission of international    
cooperation between ESA and NASA. NSO is operated by AURA, Inc., under   
cooperative agreement with the NSF.  
\end{acknowledgements} 
\appendix 
\section {Momenta calculation.~\label{app}} 
We give here a detailed description of equations and physical parameters  
used to calculate the downward  
and upward momenta $P_{down}$ and $P_{up}$.  
 
The mass density can be written as $\rho=\mu m_{\rm p}~n_{T}$, where   
$\mu$ is the mean particle weight, $m_{\rm p}=1.67\times10^{-24}$~g is the 
proton mass, and $n_{T}$ the total particle number density.   
For a fully-ionised plasma of hydrogen with 10~\% helium, $\mu=0.61$ and   
$n_{T}=1.91~n_{\rm e}$, while for a neutral gas, $\mu=1.27$ and   
$n_{T}=1.1~n_{\rm H}$, where $n_{\rm H}$ is the hydrogen number density.   
Downflow momentum in the chromosphere (neutral gas) can then be written as:   
\begin{equation}   
P_{down} =  1.27~m_{\rm p}~1.1~n_{ch}~S_{ch}~v_{ch}~\Delta h~f_{ch},   
\label{eq:momch}   
\end{equation}   
\noindent where $n_{ch}$ is the {\it pre-flare} chromospheric hydrogen   
density, assumed to be (2.2~to~4.0)$\times10^{13}$~cm$^{-3}$ from model VAL F 
\citep{VAL:81}. 
The area $S_{ch}$ with  
detectable chromospheric downflows  
is $2.5\times10^{17}$~cm$^2$. Within this area, the average downflow velocity, 
$v_{ch}$, is about ($7\pm3$)~km~s$^{-1}$. 
$\Delta h$ represents the thickness of the   
chromospheric condensation. We assume that $\Delta~h$ between a lower   
limit of $\approx100$~km, given by dynamic simulations    
\citep{Abbett&Hawley:99}    
and an upper limit of 200~km, obtained in semi-empirical models of a small   
flare that includes the velocity fields \citep{Falchi:02}.   
Finally, $f_{ch}$ is the chromospheric filling factor. Since the spatial   
scale of the H$\alpha$ images is $0.5\arcsec$ per pixel, smaller than the   
size of the velocity patches in the chromosphere ($2\arcsec$~to~$3\arcsec$),   
we assume $f_{ch}=1$. With these values, we obtain   
$5.1\times~10^{19}$~g~cm~s$^{-1}~\le~P_{down}~\le~4.6\times~10^{20}$~g~cm~s$^{-1}$. 
The lower and upper limits refer to the minimum   
and maximum values for the hydrogen density ($n_{ch}$), $v_{ch}$, and 
$\Delta~h$. 
   
At coronal level, the momentum of the upflowing plasma is given by:   
\begin{equation}   
P_{up} =  0.61~m_{p}~1.91~n_{\rm e}~S_{cor}~L~f~v_{cor},   
\label{eq:momtr}   
\end{equation}   
\noindent with $n_{\rm e}$ given by   
\begin{equation}   
n_{\rm e}=\sqrt{\frac{EM}{f~S_{cor}~L}}.   
\label{eq:ne}   
\end{equation}   
\noindent In the above equations $S_{cor}$ is the area of the region from   
where blue-shifted profiles   
arise, $L$ the thickness of the emitting region, $f$ the coronal   
filling factor, $v_{cor}$ the plasma   
speed   
at coronal level, and $EM$ the emission measure computed from the observed   
radiance   
using the CHIANTI database \citep{Dere-etal:97,Landi-etal:99} with   
ionisation equilibrium calculations from \citet{Mazzotta-etal:98}.   
 
The radiance of the Fe~{\sc xix}~59.2~nm line is proportional to   
${n_{\rm e}}^2$ for $8~\le~\log~n_{\rm e}/{\rm cm^{-3}}~\le~12$ and it is, 
hence, suitable for   
emission measure calculation.   
The iron abundance $A_{[\rm Fe]}$ is 8.1 in the corona 
\citep{Feldman-etal:92}. However, because the 
plasma is ablated from the chromosphere, the photospheric abundance of 7.6 
\citep{Grevesse-Sauval:98} may also apply. Thus, we will assume an abundance 
varying between 7.5 and 8.1. 
 
Coronal upflows are measured in eleven effective CDS pixels, so    
$S_{cor}=1.2\times~10^{18}$~cm$^2$. $L$ is taken to vary between   
1.1$\times10^{9}$~cm 
($=\sqrt{S_{cor}}$) and 2$\times10^{9}$~cm 
(obtained estimating the geometrical depth of the flaring   
loop along the line of sight, assuming the loop is perpendicular to the   
solar surface). 
The electron density, $n_{\rm e}$, of the evaporating plasma   
(from the emission measure) and the coronal velocity, $v_{cor}$, depend on whether   
the Fe~{\sc xix} spectra are fitted with one or two components.   
The computed values of $n_{\rm e}$ are between $3\times10^9$ and  
$8\times 10^9$~cm$^{-3}$ 
for $f$=1. 
%Since the spatial scale of our   
%CDS observations is $\approx 80$ times worse than that of the ground data,   
%we left $f$ free to vary.  
Finally, we note that in the extreme case of magnetic flux   
tubes not expanding with height, the ratio $S_{ch}/S_{cor}$ yields a lower 
limit of $f=0.2$, so we assume $0.2~\leq~f~\leq~1$. 
 
For both values of $f$, $P_{up}$ is   
calculated with the values of $EM$ and $v_{cor}$ obtained in both cases of   
single and double component fitting and for both estimates of $L$ and 
$A_{[\rm Fe]}$. 
The maximum and minimum of the sixteen obtained values are taken to give 
$5.9\times~10^{19}$~g~cm~s$^{-1}~\le~P_{up}~\le~4.9\times~10^{20}$~g~cm~s$^{-1}$. 
   
%%%%%%%%%%%%%%%%%%%%%%%%%%%%%%%%%%%%%%%%%%%%%%%%%%%%%%%%%%%%%%%%%%%%%%%%%%%%%%%   
%%%%%%|   
%%%%%%|  REFERENCES   
%%%%%%|   
%%%%%%%%%%%%%%%%%%%%%%%%%%%%%%%%%%%%%%%%%%%%%%%%%%%%%%%%%%%%%%%%%%%%%%%%%%%%%%%   
   
\bibliographystyle{aa}   
   
\bibliography{5065}
   
\end{document}